\documentstyle[11pt,paspconf,epsf]{article}

\begin{document}
\title{Small-Scale Mixing of the Elements}
\author{Bruce G. Elmegreen}
\affil{IBM Research Division, T.J. Watson Research Center,
P.O. Box 218, Yorktown Heights, NY 10598 USA, bge@watson.ibm.com}

\begin{abstract}
The progressive mixing and contamination of interstellar gas
by supernovae and other processes following the passage of a spiral
density wave is reviewed, with an emphasis on the Solar neighborhood.
Regions of star formation should begin their lives with an
inhomogeneous mixture of abundances as a result of their chaotic and
large-scale formation processes.  These inhomogeneities should
continue to increase during several generations of star formation
until the gas enters the interarm region.  Then cloud dispersal by
large interarm tidal forces, high rates of shear, and internal star
formation should all lead to cloud-to-cloud mixing, while increased
ionization, heating, and evaporation should lead to mixing on the
atomic level.  Gradients in the dispersion of elemental abundances are
expected for galaxy disks.
\end{abstract}


{\it to be published in} "Abundance Profiles: Diagnostic Tools
for Galaxy History," ed. D. Friedli, M.G. Edmunds, C. Robert, 
\& L. Drissen, ASP Conference Series, 1998.

\section{Introduction: The Star-Formation Process}

Much of the interstellar gas is hierarchically structured, with
relative velocities that increase approximately as the square root of
the separation.  These and other power law correlations between cloud
properties imply that the structure is scale-free, in which case the
title of this paper, "small-scale mixing" can actually refer to a wide
range of scales.  We restrict the discussion, though, to contamination
and mixing processes that operate on scales smaller than or 
equal to one scale height.  This includes much of the gas dynamics
prior to star formation, essentially all of it during star formation,
and some of it after star formation.  We also consider only gas
processes, and not stellar migration or other stellar mixing
processes, even though these stellar processes may dominate the
observed star-to-star abundance dispersion at any particular galactic
radius.  A third restriction is to galaxies like the Milky Way, and in
some discussions, to the Solar neighborhood of the Milky Way, because
this is the type of region where we know the most about abundance
dispersions.

\section{Mixing of the ISM Before Star Formation Begins}

\begin{figure}[t]
\plotfiddle{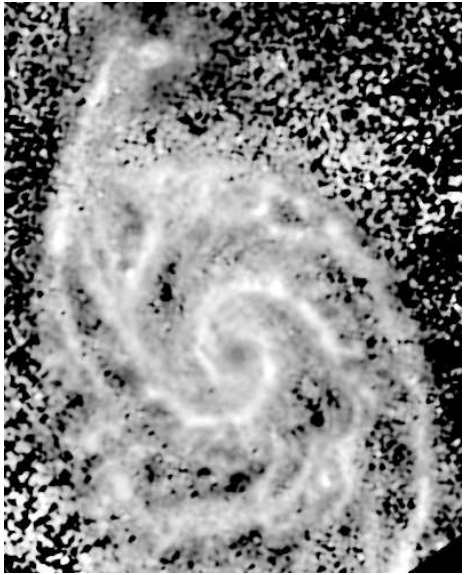}{70truemm}{0}{40}{40}{-130}{-50}
\caption{An image of M51 made from
the ratio of 15$\mu$ emission to B band emission, showing both
the warm emitting dust and the cold absorbing dust as bright features.
The gas concentrations that lead to star formation in the spiral arms
are easily seen at this resolution of 15'.  Image from Block et al.
(1997).  }
\label{fig:m51}
\end{figure}

Most star formation seems to begin on a large scale, comparable to the
thickness of the Galaxy disk.  In galaxies like ours, which contain
strong spiral density waves, it begins in the spiral arms in giant gas
clouds that typically have a mass of $\sim10^7$ M$_\odot$.  These are
observed in atomic gas in many galaxies, such as the Milky Way
(Grabelsky et al.  1987; Elmegreen \& Elmegreen 1987), M100 (Knapen et
al.  1993), and M31 (Lada et al.  1988), and they are observed in
molecular gas in other galaxies, such as M51 (Rand \& Kulkarni 1990).

Figure \ref{fig:m51} shows the gas distribution for M51 made from a
combination of dust absorption in the B band and dust emission at 15
$\mu$ from ISO (Block et al.  1997).  The large knots in the spiral
arms, separated by several kiloparsecs, are the main regions of star
formation, as shown by H$\alpha$ and CO.  Several giant shells can
also be seen in the northern and southern arms in the region of
corotation.

Spiral arm gas concentrations like those shown in figure \ref{fig:m51}
presumably form by the gravitational collapse of
density-wave-compressed gas, involving collapse motions along the
direction of the arm.  The existing theory of this process gives the
correct cloud spacing of $\sim3 $ spiral arm widths, or about
2.5 kpc in galaxies like ours, and it gives the correct
cloud mass and a short enough time scale for these clouds to form
while the gas is still in the arm (Elmegreen 1994).  The formation of
such giant clouds is not limited to density wave arms, however, since
equally large clouds appear in galaxies with only weak arms
(Thornley \& Mundy 1997a,b) 
and in irregular galaxies (Cohen et al. 1988; Hunter 1997).  The
peculiar conditions in the crest of a strong spiral arm, namely, the
high gas density, the low rate of shear, and the low galactic tidal
force, make this place highly favored over other sites in galaxies
that have spiral waves.

The onset of star formation in spiral arms implies that star-forming
clouds begin their lives in violent conditions, possibly inside shock
fronts, over a galactic radial range of $\sim1$ kpc, considering a
typical arm pitch angle of $\sim15^\circ$.  This means that the
pre-star-forming gas will be highly turbulent, and it will have a
range of metallicities from the background galactic gradient that may
be $\pm0.05$ dex.

\begin{figure}[t]
\plotfiddle{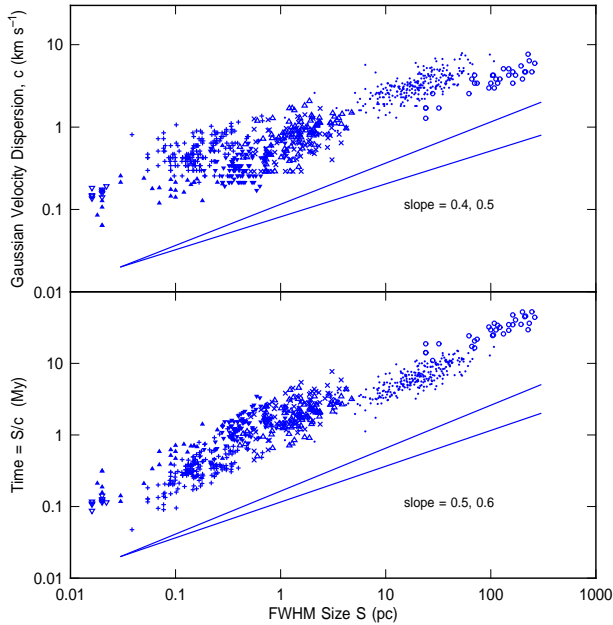}{80truemm}{0}{100}{100}{-300}{-260}
\caption{Size and linewidth data from GMC surveys. Symbols are for GMC
surveys = {\it dots}: Solomon et al. (1987), {\it open circles}: Dame et al. (1986);
quiescent clouds = {\it filled triangles}: Falgarone et al. (1992), {\it open
triangles}: Williams et al. (1994; the Maddalena-Thaddeus cloud),
{\it inverted open triangles}: Lemme et al. (1995; L1498),
{\it inverted filled triangles}: Loren (1989; Ophiuchus),
OB associations = {\it crosses}: Williams et al. (1994; Rosette),
{\it plus signs}: Stutzki \& G\"usten (1990; M17).}
\label{fig:sl}
\end{figure}

\begin{figure}[t]
\plotfiddle{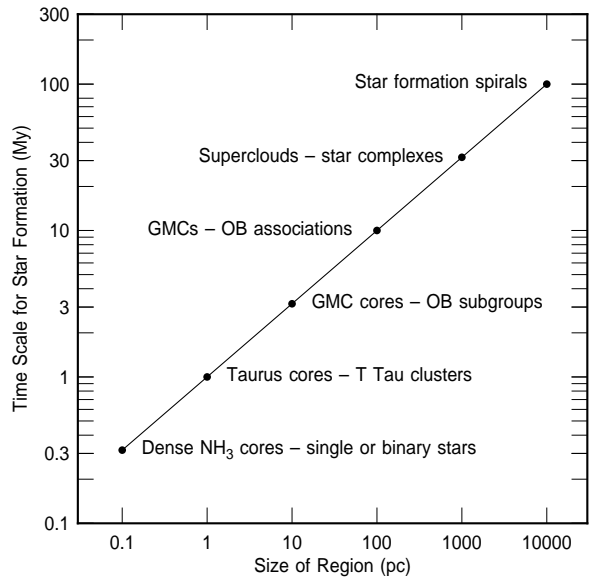}{70truemm}{0}{100}{100}{-280}{-310}
\caption{Schematic diagram showing how the duration of star formation
in regions of various sizes increases with the square root of the
size.}
\label{fig:hierarchies}
\end{figure}

The onset of star formation inside such a cloud is relatively quick.
It seems to being after a time comparable to the turbulent crossing
time, regardless of scale.  This time is is $\sim0.7S(pc)^{0.5}$ My
for scale size $S$, as determined from the molecular cloud
size-linewidth correlation, shown in figure \ref{fig:sl}.  At the top
of the figure is the linewidth versus size, and at the bottom is the
ratio of size to linewidth, versus size.  The {\it duration} of star
formation in a region is also proportional to the local crossing time,
perhaps equal to several times this for a wide range of scales
(Elmegreen \& Efremov 1996).  That is, small regions or small clumps
form their stars quickly and disperse, while large regions or large
clouds take a long time.  A schematic diagram of this trend is shown
in Figure \ref{fig:hierarchies}.  The correlation between crossing
time and size, shown at the bottom of figure \ref{fig:sl}, is 
similar to the correlation between duration of star formation and
size, shown in figure \ref{fig:hierarchies}.

A quick onset to star formation in a cloud born under violent and
turbulent conditions implies that any non-uniformity of abundances
that may enter the cloud, from an initial non-uniformity in the
pre-cloud gas, for example, or from a galactic gradient, will not
smooth out much before star formation actually begins.	That is, {\it
there is not much time for mixing before star formation.} Indeed, the
dominant mode of star formation in galaxy disks involves giant cloud
complexes that are likely to be initially inhomogeneous in elemental
distribution at a level of perhaps $\pm0.05$ dex.  The stars that form
in these clouds should be equally inhomogeneous.

\section{Mixing of Supernova Debris During Star Formation}

Hierarchical structure in star-forming regions implies that star
formation is concentrated into a small fraction of the volume, and the
ensuing supernova contamination is localized.  For fractal clouds, the
volume filling factor of dense gas is \begin{equation}f_{\rm
dense\;\;gas}={\rm peak\;density\;contrast}^{(D/3)-1} \sim0.1
\end{equation} for fractal dimension $D=2.3$ (Elmegreen \& Falgarone
1996) and peak density contrast of $\sim10^4$.	Thus, in any region,
star formation occurs in only 10\% of the volume.

The typical size of a supernova remnant in a region of star formation
is only $\sim10-50$ pc.  From Cioffi, McKee \& Bertschinger (1988) the
final size of a remnant following the pressure-driven snowplow phase
is
\begin{equation}
R_{final}=4.93R_{PDS}\left({{E_{51}^{1/14}n_0^{1/7}}\over{a_{10}}}\right)^{3/7}
\end{equation}
where
\begin{equation}
R_{PDS}(pc)=14.0\left({{E_{51}^{2/7}}\over{n_0^{3/7}}}\right).
\end{equation}
Here $E_{51}$ is the supernova energy, $n_0$ is the preshock density,
and $a_{10}$ is the preshock velocity dispersion in units of $10$ km
s$^{-1}$.  Thus
\begin{equation}
R_{final}(pc)=69E_{51}^{0.32}n_0^{-0.37}a_{10}^{-0.43}.
\end{equation}
But $a({km\;s^{-1}})\sim0.7S(pc)^{1/2}$ from the molecular cloud
correlations, and $n_0\sim5000S(pc)^{-1}$ for pressure $\rho
a^2\sim4\times10^5$ k$_B$ in a star-forming region.  Then we get
\begin{equation}
R_{final}(pc)\sim10E_{51}^{0.32}S(pc)^{0.15},
\end{equation}
which is relatively insensitive to scale.  If the supernova occurs in
a region that is cleared of gas by previous HII regions and wind
pressures, then it can become larger than this.

The size of a region of star formation in which OB stars form is
typically larger than the size of a supernova remnant in such a
region.  There {\it is} a characteristic (or minimum) size for such OB
star-forming regions even though there is no characteristic size for
star-forming regions in general.  This is because it takes a minimum
number of $\sim10^4$ stars before an O-type star is likely to form at
all under a random sampling of the initial stellar mass function.
Thus low mass clouds tend to form only low mass stars (because these
stars are common) and high mass clouds tend to form OB stars in
addition to low mass stars.  The star-forming regions in which
supernovae are likely to occur span a distance scale greater than
$\sim10$ pc, possibly even greater than $\sim 100$ pc, as suggested
also by the plotted points for OB subgroups and associations in figure
\ref{fig:sl}.  These scales are comparable to or larger than the
associated supernova remnant sizes, so {\it supernova contamination in
a region of star formation should be spotty and localized}.  The
supernova debris will therefore mix with only part of the star-forming
gas, and may even get highly concentrated in several small regions if
the explosion happens to occur at close range to the cloud.  Note that
$R_{final}$ in the above equation is small, so according to this, a
lot of contamination will be at a range less than $\sim10$ pc.

Debris from multiple supernovae in an OB association should
contaminate a few dense clouds and leave the rest in a shadow or far
away.  Thus {\it the clumpy, hierarchical nature of star-forming
regions should lead to abundance inhomogeneities during the star
formation/supernova epoch}, which is a $3-20$ My year period following
the onset of star formation in a region where OB stars form.  Studies
of supernova remnants in regions of star formation were made by Junkes
et al.	(1992) and others.

\section{Penetration of Supernova Debris into a Shocked Cloud}

In a simple bow shock, there is a contact discontinuity between the
dense cloud and the SN gas, and this surface has no direct mixing in
equilibrium conditions.  Kelvin-Helmholtz and Rayleigh-Taylor
instabilities on this surface will mix the gas, however, and these
instabilities should be responsible for contamination of the cloud
by supernova material.	In numerical simulations of this process, such
mixing is invariable {\it behind the cloud}, and not in the cloud core
(e.g., Xu \& Stone 1995).

It is unknown if SN debris also mixes with the gas in a compressed
cloud core.  Stone \& Norman (1992) simulate a shock/cloud collision
with $\gamma=5/3$ and suggest that a Richtmyer-Meshkov instability
might lead to mixing at the front surface and inside the core.	Foster
\& Boss (1996, 1997) consider a shock/cloud collision with $\gamma=1$
and get a very thin region between the cloud and the shock in the SN
ejecta.  This thin region allows Rayleigh-Taylor fingers at the head
of the cloud to penetrate into the core and aid mixing with dense,
pre-star-formation material.

Supernova interactions with fractal or hierarchically structured clouds
have not yet been modeled.

\section{Mixing of SN-contaminated Clouds after SF}

The size-linewidth-time relation in a hierarchical cloud implies that
many small regions of SF come and go before the large region is
finished.  This is also evident from the schematic diagram in figure
\ref{fig:hierarchies}.	One implication of this is that the large
star-forming clouds, which cannot be easily destroyed by internal star
formation, probably move around and get recycled as even larger scales
in the hierarchy continue to form stars.  In this case, GMC's can have
several epochs of supernova contamination, and each one can increase
the dispersion of the final stellar abundances in a stochastic fashion
before all star formation stops in the region.

An example is offered by the formation in the Solar neighborhood.
Most of it began 30-50 My ago when the neighborhood was shocked by the
local spiral arm, which is now the Carina arm.	This assumes a pattern
speed for the Galactic spiral system of 13.5 km s$^{-1}$ kpc$^{-1}$
from Yuan (1969).  The first generation of star formation 
probably made the
Cas-Tau association (Blaauw 1984), in addition to some
clusters, such as $\zeta$ Sculptoris and
the Pleiades.
This first generation also could have caused the expansion of
Lindblad's ring, which now seems to contain the Orion, Perseus, and
Sco-Cen OB associations along the periphery (see review in P\"oppel
1997).	The Cas-Tau association at the center of this ring now has
very little gas, so presumably its residual gas and other spiral arm
gas got collected into
Lindblad's ring, and eventually into the local associations.  Thus the
current supernova contamination of Orion and other local clouds is
probably the second time this has happened since the spiral wave
passed by.

The tilt of Gould's Belt has been attributed to the impact of an
extragalactic cloud (Franco et al. 1988; Comer\'on \& Torra 1993), 
but there are other explanations
such as a Parker instability (Gomez de Castro \& 
Pudritz 1992; Shibata \& Matsumoto 1991), and
general perturbations from the LMC (Edelsohn \& Elmegreen 1997).
Maybe the spiral arm itself is responsible for the tilt.  The shock in
a spiral arm contains more than enough speed to elevate the 
gas to one scale height; any asymmetry in the spiral potential
relative to the gas midplane could conceivable lead to a tilted gas
distribution when giant clouds form.

The primary disadvantages of the cloud impact model are (1) that the
local star formation seemed to begin just when the local region was at
or near the crest of a spiral arm, and such timing for an impact would
be unlikely, and (2) that corrugations like this are common in the
spiral arms of the inner galaxy (Pandey, Bhatt, \& Mhara 1988; Sanders, 
Solomon,
\& Scoville 1984) and other galaxies (Florido et al. 1991), 
and it is unlikely all were
formed by cloud impacts.  Nevertheless, the impact model
is intriguing from the point of view of abundance anomalies
in Orion and other local regions, because an impacting high velocity
cloud could have contaminated the disk with low metallicity material
(Cunha \& Lambert 1992; Edvardsson et al. 1995).

\section{Mixing of SN-contaminated Cloud Debris after SF}

Ultimately the star-forming gas gets converted to low density by
ionization, heating, and evaporation.  These processes cause {\it
atomic level} mixing.

The large-scale star-forming regions are also stretched by increased
shear and tidal forces when they enter the interarm.  These processes
cause {\it cloud-to-cloud} mixing rather than atom-to-atom mixing, but
they also expose the dense gas to stray stellar enegy which can lead
to more ionization, heating, and evaporation.  It follows that {\it
passing from an arm to the interarm region should help homogenize
the gas at both the cloud and the atomic level.}

In the inner part of the Milky Way, the next spiral density wave arm
arrives rather quickly, in only 50 My years or less.  This is before
cloud destruction is complete so there is a significant amount of
dense molecular gas and star formation still in the interarm.
For example, figure \ref{fig:m51} shows a lot more gas and dense
knotty emission in the interarm region at small radii than large
radii.

Based on such analyses, one might expect cloud self-destruction to be
more complete in the interarm regions in the outer part of the Milky
Way than the inner part.  This implies that the interarm
homogenization of supernova enriched gas should change with radius,
and so should the relative dispersion in abundances.

Post-star-formation mixing is also possible in chimney vents and giant
bubbles that send enriched material into the halo where it can
disperse over a radial range of perhaps a kiloparsec or more.  In the
hot gas, the supernova debris can mix atom-by-atom, so this form of
dispersal is at the atomic level.

\begin{figure}[t]
\plotfiddle{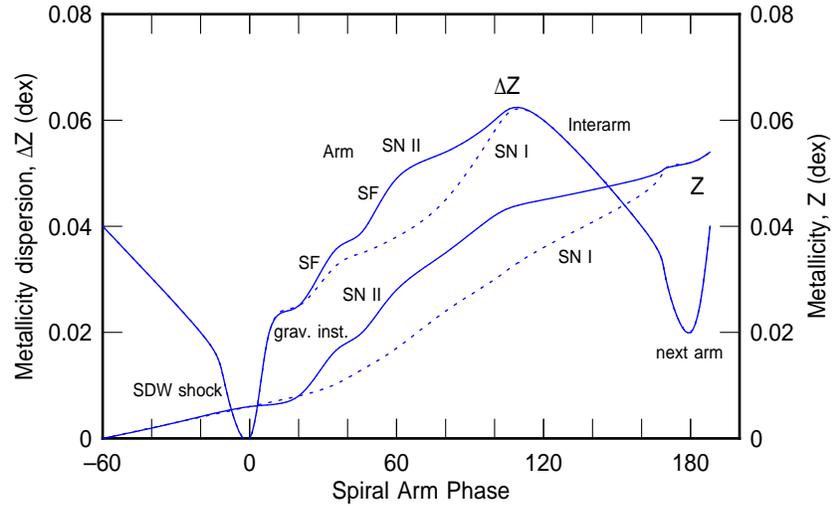}{60truemm}{0}{100}{100}{-160}{-130}
\caption{Schematic diagram of the change with spiral arm phase of the
elemental abundance and the dispersion in the elemental abundance,
based on the qualitative model presented in the text.
Features related to star formation (SF), spiral arms and interarms, 
spiral density wave (SDW) shocks, and supernovae (SN) are shown. 
}
\label{fig:zscheme} \end{figure}

\section{Conclusions}

1. Pre-star formation clouds should be inhomogeneous as a result of
their chaotic and large-scale formation processes.\\
2.  Inhomogeneities should increase during star formation as a result
of the clumpy, hierarchical structure of young stars.  The penetration
depth of SN debris into triggered clouds is unknown, however.\\
3.  Short-time cloud recycling should continue to increase the
metallicity and the metallicity dispersion until all star formation
stops.\\
4.  Ionization, heating, evaporation, and hot chimneys should
homogenize the gas at the atomic level, while general galactic shear
should mix at the cloud-to-cloud level.\\
5.  Passing to an interarm region should disperse a cloud complex and
aid the homogenization process by allowing deeper penetration of
background radiation.

A schematic diagram of the progressive change in abundance and
abundance dispersion with phase in a spiral density wave is shown in
figure \ref{fig:zscheme}.  The contaminations from Type I and Type II
supernova debris are tracked separately, because only the Type II
debris (e.g., Oxygen) correlates with star formation.  The Type I
debris (e.g.  Iron) is added more continuously to the gas as the comoving
old stars evolve into supernovae.  The
vertical scales on the left and right are pure conjecture, but they
sensibly follow from the discussion given above.

\end{document}